\documentclass[a4paper, oneside, twocolumn, notitlepage, 10pt]{extarticle_ecoc}
\usepackage{ecoc}

\usepackage[acronym]{glossaries}
\newacronym{acf}{ACF}{autocorrelation function}
\newacronym{air}{AIR}{achievable information rate}
\newacronym{app}{APP}{a posteriori probability}
\newacronym{ase}{ASE}{amplified spontaneous emission}
\newacronym{ask}{ASK}{amplitude-shift keying}
\newacronym{awgn}{AWGN}{additive white Gaussian noise}

\newacronym{b2b}{B2B}{back-to-back}
\newacronym{baa}{BAA}{Blahut-Arimoto Algorithm}
\newacronym{bpcu}{bpcu}{bits per channel use}

\newacronym{cd}{CD}{chromatic dispersion}
\newacronym{cdf}{cdf}{cumulative density function}
\newacronym{coi}{COI}{channel of interest}
\newacronym{cpan}{CPAN}{correlated phase and additive noise}
\newacronym{cscg}{CSCG}{circularly symmetric complex Gaussian}
\newacronym{csi}{CSI}{channel state information}
\newacronym{cwt}{CWT}{continouus-wave tone}

\newacronym{dbp}{DBP}{digital backpropagation}
\newacronym{dsp}{DSP}{digital signal processing}
\newacronym{dtft}{DTFT}{discrete time Fourier transform}

\newacronym{dd}{DD}{direct detection}
\newacronym{dmd}{DMD}{differential mode delay}
\newacronym{dmgd}{DMGD}{differential mode group delay}
\newacronym{dgd}{DGD}{differential group delay}
\newacronym{dt}{DT}{decision tree}

\newacronym{evd}{EVD}{Eigenvalue decomposition}

\newacronym{fmf}{FMF}{few-mode fiber}
\newacronym{fwhm}{FWHM}{full-width half-maximum}
\newacronym{fwm}{FWM}{four-wave mixing}

\newacronym{ghq}{GHQ}{Gauss-Hermite-Quadrature}

\newacronym{idra}{IDRA}{ideal distributed Raman amplification}
\newacronym{iid}{i.i.d.}{independent and identically distributed}
\newacronym{im}{IM}{intensity modulation}
\newacronym{isi}{ISI}{inter-symbol interference}
\newacronym{iss}{ISS}{inter-symbol sample}

\newacronym{jd}{JD}{joint detection}
\newacronym{jdd}{JDD}{joint detection and decoding}
\newacronym{jdabaa}{DABAA}{dynamic-assignment Blahut-Arimoto Algorithm}

\newacronym{kld}{KLD}{Kullback-Leibler-Divergence}
\newacronym{kk}{KK}{Kramers-Kronig}

\newacronym{lstm}{LSTM}{long short-term memory}
\newacronym{lp}{LP}{linearly polarized}

\newacronym{mcf}{MCF}{multi-core fiber}
\newacronym{md}{MD}{modal dispersion}
\newacronym{mgdm}{MGDM}{mode-group-division multiplexing}
\newacronym{mdg}{MDG}{mode-dependent gain}
\newacronym{mdl}{MDL}{mode-dependent loss}
\newacronym{mdm}{MDM}{mode-division multiplexing}
\newacronym{me}{ME}{Manakov equation}
\newacronym{mi}{MI}{mutual information}
\newacronym{mir}{MIR}{mutual information rate}
\newacronym{mimo}{MIMO}{multiple-in multiple-out}
\newacronym{miso}{MISO}{multiple-in single-out}
\newacronym{ml}{ML}{machine learning}
\newacronym{mmf}{MMF}{multi-mode fiber}
\newacronym{mpc}{MPC}{minimum-phase criterion}
\newacronym{mzm}{MZM}{Mach-Zehnder modulator}

\newacronym{nlse}{NLSE}{nonlinear Schrödinger equation}
\newacronym{nn}{NN}{neural network}

\newacronym{pam}{PAM}{pulse-amplitude modulation}
\newacronym{pdf}{pdf}{probability density function}
\newacronym{pmf}{pmf}{probability mass function}
\newacronym{psd}{PSD}{power spectral density}
\newacronym{psk}{PSK}{phase-shift keying}

\newacronym{qam}{QAM}{quadrature-amplitude modulation}

\newacronym{roadm}{ROADM}{reconfigurable optical add-drop multiplexer}
\newacronym{rp}{RP}{regular perturbation}
\newacronym{rrc}{RRC}{root-raised cosine}
\newacronym{rv}{RV}{random variable}

\newacronym{sd}{SD}{seperate detection}
\newacronym{sdd}{SDD}{separate detection and decoding}
\newacronym{sdm}{SDM}{space-division multiplexing}
\newacronym{sic}{SIC}{successive interference cancellation}
\newacronym{simo}{SIMO}{single-in multiple-out}
\newacronym{sinr}{SINR}{signal to interference and noise ratio}
\newacronym{siso}{SISO}{single-in single-out}
\newacronym{sld}{SLD}{square-law detector}
\newacronym{snr}{SNR}{signal-to-noise ratio}
\newacronym{smf}{SMF}{single-mode fiber}
\newacronym{spa}{SPA}{sum-product algorithm}
\newacronym{spm}{SPM}{self-phase modulation}
\newacronym{ssb}{SSB}{single side-band}
\newacronym{ssbi}{SSBI}{signal-signal beat interference}
\newacronym{ssmf}{SSMF}{standard single-mode fiber}
\newacronym{svd}{SVD}{singular value decomposition}
\newacronym{svm}{SVM}{state vector machine}

\newacronym{te}{TE}{transverse electric}
\newacronym{tm}{TM}{transverse magnetic}

\newacronym{urr}{URR}{unidistant Rayleigh ring}

\newacronym{wdm}{WDM}{wavelength-division multiplexing}
\newacronym{wlg}{w.l.g.}{without loss of generality}

\newacronym{xpm}{XPM}{cross-phase modulation}

\usepackage{xcolor}

\usepackage{multicol}

\usepackage{amsmath}
\usepackage{amssymb}
\usepackage{bm} %

\usepackage{tikz}
\usepackage{tikz-3dplot}
\usetikzlibrary{shapes.multipart}
\usetikzlibrary{positioning}
\usetikzlibrary{arrows}
\usetikzlibrary{shapes.multipart}
\usetikzlibrary{shapes.geometric,shapes.arrows,decorations.pathmorphing}
\usetikzlibrary{matrix,chains,scopes,positioning,arrows,fit}
\usetikzlibrary{shapes.multipart}
\usetikzlibrary{calc}
\usetikzlibrary{decorations.markings}
\usepackage{pgfplots}
\pgfplotsset{compat=1.18}
\usetikzlibrary{arrows.meta,arrows}

\usepackage[detect-all]{siunitx}
\DeclareSIUnit\bpcu{\gls{bpcu}}

\usepackage{subcaption}
\usepackage{hyperref}

\newcommand{\figref}[1]{Fig.~\ref{#1}}

\newcommand{\vek}[1]{{\boldsymbol{#1}}}

\renewcommand{\j}{{\mathrm{j}}}
\newcommand{\e}{{\mathrm{e}}}

\newcommand{\nsym}{\ensuremath{n}}

\newcommand{\incmean}{\ensuremath{\mu_\delta}}
\newcommand{\incvar}{\ensuremath{\sigma_\delta^2}}
\newcommand{\incstd}{\ensuremath{\sigma_\delta}}
\newcommand{\thetavar}{\ensuremath{\sigma_\theta^2}}

\newcommand{\nphases}{{\ensuremath{n_p}}}
\newcommand{\ptx}{{\ensuremath{P_\mathrm{tx}}}}

\definecolor{matlab1}{rgb}{0.00000,0.44700,0.74100}%
\definecolor{matlab2}{rgb}{0.85000,0.32500,0.09800}%
\definecolor{matlab3}{rgb}{0.92900,0.69400,0.12500}%
\definecolor{matlab4}{rgb}{0.49400,0.18400,0.55600}%
\definecolor{matlab5}{rgb}{0.46600,0.67400,0.18800}%
\definecolor{matlab6}{rgb}{0.30100,0.74500,0.93300}%
\definecolor{matlab7}{rgb}{0.63500,0.07800,0.18400}%
\definecolor{matlabBlue}{rgb}{0.00000,0.44700,0.74100}%
\definecolor{matlabRed}{rgb}{0.85000,0.32500,0.09800}%

\begin{document}
\selectlanguage{english}    %

\title{Successive Interference Cancellation for Optical Fiber Using Discrete Constellations}%

\author{
    Alex Jäger, Gerhard Kramer
}

\maketitle                  %

\begin{strip}
    \begin{author_descr}

        Institute for Communications Engineering, School of Computation, Information and Technology, Technical University of Munich, 80333, Munich, Germany,
        \textcolor{blue}{\uline{alex.jaeger@tum.de}, \uline{gerhard.kramer@tum.de}}

    \end{author_descr}
\end{strip}

\renewcommand\footnotemark{}
\renewcommand\footnoterule{}

\begin{strip}
    \begin{ecoc_abstract}
        Successive interference cancellation is used to detect discrete modulation symbols transmitted over a 1000 km fiber-optic link. A transmitter and receiver are presented that have linear complexity in the number of transmitted symbols and achieve the information rates of previous studies that use continuous modulations. \textcopyright2024 The Author(s)
    \end{ecoc_abstract}
\end{strip}
\glsresetall
\section{Introduction}
We investigate communication via \gls{wdm} through optical networks where receivers can access only their own \gls{wdm} channel. Motivated by first-order regular perturbation \cite{Vannucci:02:RP,Mecozzi:12:RP}, the distortions caused by \gls{xpm} are modeled as phase noise with correlations over many symbols \cite{Dar:13:NLIN,Dar:16:Collision}. 
Lower bounds on the end-to-end \gls{mi} after \gls{jdd} were developed in \cite{Secondini:19:Wiener,Gomez:20:CPAN} by using particle filtering. However, it is not obvious how to combine such receiver algorithms with concrete coded modulations. For example, turbo detection and decoding \cite{douillard1995iterative,Colavolpe:05:Iterative,Yankov:15:Phase} is an option but requires dedicated code design and receiver iterations\cite{ten2004design}. Instead, a receiver that uses \gls{sic} allows using off-the-shelf codes\cite{jaeger:24:sic}; see also\cite{Wachsmann:99:Multilevel,Pfister:01:ISI,Plabst:24:SIC}. Prior work studied receiver performance for continuous alphabet signaling. In this contribution, we investigate discrete modulation using probabilistically-shaped star \gls{qam}.

\section{System Model}
Consider transmitting a vector $\vek{x}$ of $\nsym$ symbols sampled independently from a constellation $\mathcal{X}$. After pulse-shaping with a sinc-filter, the signal propagates over a channel described by the \gls{nlse}. The signal is disturbed by \gls{wdm} signals co-propagating at different wavelengths, and the receiver can access only the wavelengths of interest through a bandpass filter. The receiver applies sampling, single-channel \gls{dbp}, matched filtering using a sinc-filter and downsampling to symbol rate to obtain the received vector $\vek{y}$.

The receiver approximates the channel from the input $\vek{x}$ to the output $\vek{y}$ using a \gls{cpan} model \cite{Gomez:20:CPAN}:
\begin{equation}
    y_i \approx x_i\e^{\j\theta_i}+n_i
\end{equation}
where the phase noise is modeled as a Gaussian process $\theta_i = \incmean \theta_{i-1}+\incstd \delta_i$ where $\delta_i$ is sampled independent of $\theta_{i-1}$ and $\vek{x}$ from a standard Gaussian distribution. The values $\incmean$ and $\incstd$ are chosen so the steady-state variance $\thetavar=\incstd^2/(1-\incmean^2)$ has a desired value which can be found by training \cite{Gomez:20:CPAN}. The additive noise $n_i$ is sampled from a \gls{cscg} distribution with variance $\sigma_n^2$. We thus have the complex-alphabet surrogate channel model
\begin{equation}
    q(y_i|x_i,\theta_i) = \frac{1}{\pi\sigma_n^2}\exp\left(-\frac{\left|y_i-x_i\e^{\j\theta_i}\right|^2}{\sigma_n^2}\right)
\end{equation}
and the conditional \gls{pdf}
\begin{equation}
    p(\theta_i|\theta_{i-1}) %
    = \mathcal{N}(\theta_i;\incmean\theta_{i-1},\incvar)
\end{equation}
where $\mathcal{N}(x;\mu,\sigma^2)$ is a Gaussian \gls{pdf} in $x$ with mean $\mu$ and variance $\sigma^2$.

\section{Constellation Design}
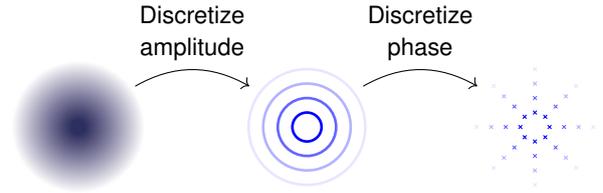
\begin{figure}
    \centering
    \begin{tikzpicture}[every text node part/.style={align=center}]
        \begin{axis}[%
            width=3.5cm,
            height=3.5cm,
            at={(-3cm,0)},
            axis lines=none,
            xmin=-10,
            xmax=10,
            ticks = none,
            ymin=-10,
            ymax=10,
            ]
            \addplot[] graphics[xmin=-10,ymin=-10,xmax=10,ymax=10] {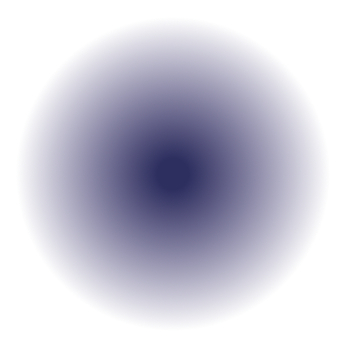};
        \end{axis}
        \begin{axis}[%
            width=3.5cm,
            height=3.5cm,
            at={(0,0)},
            axis lines=none,
            xmin=-10,
            xmax=10,
            ticks = none,
            ymin=-10,
            ymax=10,
            ]
            
            \draw[line width = 1pt, blue] (0,0) circle[radius=2];
            \draw[line width = 1pt, blue, opacity = 0.6] (0,0) circle[radius=4];
            \draw[line width = 1pt, blue, opacity = 0.3] (0,0) circle[radius=6];
            \draw[line width = 1pt, blue, opacity = 0.1] (0,0) circle[radius=8];
        \end{axis}
        \begin{axis}[%
            width=3.5cm,
            height=3.5cm,
            at={(3cm,0)},
            axis lines=none,
            xmin=-10,
            xmax=10,
            ticks = none,
            ymin=-10,
            ymax=10,
            ]    
            \addplot [color=blue, forget plot, mark=x, mark size = 1, only marks]
            table[]{data/star-qam-1.tsv};
            \addplot [color=blue, opacity = 0.6, mark=x, mark size = 1, only marks]
            table[]{data/star-qam-2.tsv};
            \addplot [color=blue, opacity = 0.3, mark=x, mark size = 1, only marks]
            table[]{data/star-qam-3.tsv};
            \addplot [color=blue, opacity = 0.1, mark=x, mark size = 1, only marks]
            table[]{data/star-qam-4.tsv};
        \end{axis}    
        \draw[->] (-1.3cm,1.5cm) to[bend left] node[midway, above] {\small Discretize\\ \small amplitude} (0.2cm,1.5cm);
        \draw[->] (1.7cm,1.5cm) to[bend left] node[midway, above] {\small Discretize\\ \small phase} (3.2cm,1.5cm);
       \end{tikzpicture}
    \caption{Discretizing a \gls{cscg} density to obtain a shaped star \gls{qam} constellation. Brightness indicates a-priori probability.}
    \label{fig:constellation}
\end{figure}

The receiver design in \cite{jaeger:24:sic} requires the conditional \gls{pdf} $q(y_i|\theta_i)=\int p(x_i)q(y_i|x_i,\theta_i)\mathrm{d}x_i$ to be constant in $\theta_i$, which is the case for \gls{cscg} and \gls{urr} distributions. The latter is a ring constellation obtained by discretizing a \gls{cscg} in amplitude.

To obtain a discrete constellation, we further discretize \gls{urr} distributions in phase to obtain probabilistically-shaped star-\gls{qam} constellations, see \figref{fig:constellation}. The absolute value $r_i$ and phase $\gamma_i$ of $x_i$ are modulated independently and have \glspl{pmf}
\begin{equation}
\label{equ:apriori}
\quad P(r_i) =  \frac{r_i}{C}\exp\left(-\frac{r_i^2}{\ptx}\right),\quad P(\gamma_i) = \frac{1}{\nphases}
\end{equation}
where $\ptx$ is the average transmit power, $C$ is a normalization constant, and $\nphases$ is the cardinality of the phase set $\left\{0,\frac{2\pi}{\nphases},2\frac{2\pi}{\nphases},\ldots,(\nphases-1)\frac{2\pi}{\nphases}\right\}$. 
For these constellations, $q(y_i|\theta_i)=\sum_{x_i} P(x_i) q(y_i|x_i,\theta_i)$ is approximately constant in $\theta_i$, and the approximation error increases in the distance of consecutive points with the same amplitude $r_i$, hence decreases in $\nphases$.

\section{Successive Interference Cancellation}
For decoding, it is practical to have independent information for each data symbol $x_i$. For example, the \gls{sdd} a-posteriori metric for $x_i$ is $q(x_i|\vek{y})$. In contrast, the \gls{jdd} a-posteriori metric $q(\vek{x}|\vek{y})$ has dependencies that give side information for each data symbol and provide improved \glspl{air} compared to \gls{sdd}. However, this metric is usually too complex for practical decoding.

\begin{figure}
    \centering
    \begin{tikzpicture}
        \node[] (y) at (0,0) {$\vek{y}$};
        \node[draw] (detectora) at (-2.7,-1) {Detector 1};
        \node[draw, below = 1.5 of detectora] (decodera) {Decoder 1};
        \node[draw] (detectorb) at (0,-1) {Detector 2};
        \node[draw, below = 1.5 of detectorb] (decoderb) {Decoder 2};
        \node[draw] (detectorc) at (2.7,-1) {Detector 3};
        \node[draw, below = 1.5 of detectorc] (decoderc) {Decoder 3};
        \node[below = 0.5 of decoderc] (output) {$\vek{r},\vek{\alpha},\vek{\beta}$};
        
        \draw[->] (y) -- (detectora);
        \draw[->] (y) -- (detectorb);
        \draw[->] (detectora) -- node[midway, fill={white}] {\small $q(r_i|\vek{y})$} (decodera);
        \draw[->] (detectorb) -- node[midway, fill={white}] {\small $q(\alpha_i|\vek{y},\vek{r})$} (decoderb);
        \draw[->] (decodera) -- node[midway, fill={white}] {\small $\vek{r}$} (detectorb);
        \draw[->] (y)--(detectorc);
        \draw[->] (decoderb)--node[midway, fill = {white}] {\small $\vek{r},\vek{\alpha}$} (detectorc);
        \draw[->] (detectorc)--node[midway, fill = {white}] {\small $q(\beta_i|\vek{y},\vek{r},\vek{\alpha})$} (decoderc);
        \draw[->] (decoderc)--(output);
    \end{tikzpicture}
    \caption{\gls{sic} with two stages.}
    \label{fig:sic}
\end{figure}
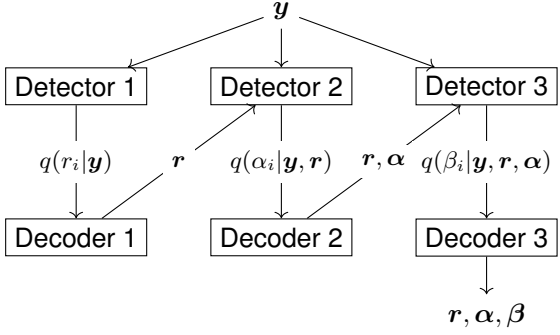

We use \gls{sic} to provide practical a-posteriori metrics while keeping the loss in terms of \gls{air} compared to \gls{jdd} small. For two \gls{sic}-stages, the phase $\vek{\gamma}=[\gamma_1,\gamma_2,\ldots]$ of the transmit vector is divided into vectors $\vek{\alpha}$ and $\vek{\beta}$ of length $n/2$ in the fashion $\vek{\gamma}=[\alpha_1,\beta_1,\alpha_2,\beta_2,\ldots]$. The algorithm operates as follows; see \figref{fig:sic}.
\begin{enumerate}
    \item Detect and decode the absolute value using \gls{sdd}. Detector 1 passes a-posteriori information $q(r_i|\vek{y})$ to decoder 1. Decoding is error-free for long error-correcting codes and transmission rates below the \gls{air}. 
    \item Decoder 1 passes $\vek{r}$ to detector 2, which computes and passes a-posteriori information $q(\alpha_i|\vek{y},\vek{r})$ to decoder 2. 
    \item Decoder 2 passes $\vek{r}$ and $\vek{\alpha}$ on to detector 3 which computes and passes a-posteriori information $q(\beta_i|\vek{y},\vek{r},\vek{\alpha})$ to decoder 2. 
\end{enumerate}

\section{Receiver Algorithm}
For the absolute values, a memoryless detector $q(r_i|y_i)$ provides the same \gls{air} as an \gls{sdd}-detector $q(r_i|\vek{y})$ and a \gls{jdd}-detector $q(\vek{r}|\vek{y})$ \cite{jaeger:24:sic}. We find that $q(r_i|\vek{y})$ is proportional to
\begin{equation}
\begin{aligned}    %
    \sum_{\gamma_i} \int\nolimits_{-\infty}^\infty P(r_i)P(\gamma_i)p(\theta_i)q(y_i|r_i,\gamma_i,\theta_i)\mathrm{d}\theta_i.
    \end{aligned}
\end{equation}
Using \eqref{equ:apriori} and
\begin{equation}
    \sum_{\gamma_i}q(y_i|r_i,\gamma_i,\theta_i)\approx \int\limits_{-\pi}^\pi q(y_i|r_i,\gamma_i',\theta_i)\mathrm{d}\gamma_i'\propto f_i(r_i)
\end{equation}
with
\begin{equation}
    f_i(r_i) = \exp\left(-\frac{r_i^2}{\sigma_n^2}\right)I_0\left(\frac{2|y_i|r_i}{\sigma_n^2}\right)
\end{equation}
gives the mismatched a-posteriori
\begin{equation}    
    q(r_i|\vek{y}) = \frac{P(r_i)f_i(r_i)}{\sum_{\tilde{r}}P(\tilde{r})f_i(\tilde{r})}
\end{equation}
where $I_0(\cdot)$ is the modified Bessel function of the first kind and zeroth order.

For the first stage of phase detection, a memoryless detector suffices \cite{jaeger:24:sic}. For odd $i$, $q(\gamma_i|\vek{y},\vek{r})$ is proportional to
\begin{equation}
\begin{aligned}    %
    \int\limits_{-\infty}^\infty P(r_i)P(\gamma_i)p(\theta_i)q(y_i|r_i,\gamma_i,\theta_i)\mathrm{d}\theta_i.
\end{aligned}
\end{equation}
The \gls{pdf} $q(y_i|r_i,\gamma_i,\theta_i)$ is approximately proportional to the Gaussian \gls{pdf}
\begin{align}
    \mathcal{N}\left(\theta_i;m(\angle_{y_i}-\gamma_i),\frac{\sigma_n^2}{2|y_i|r_i}\right)
\end{align}
and using $p(\theta_i)=\mathcal{N}(\theta_i;0,\thetavar)$, we have
\begin{align}
    & q(\gamma_i|\vek{y},\vek{r}) = \frac{g_i(\gamma_i)}{\sum_{\tilde{\gamma}}g_i(\tilde{\gamma})}\\
    & g_i(\gamma_i) = \mathcal{N}\left(m(\angle_{y_i}-\gamma_i);0,\sigma_\theta^2+\frac{\sigma_n^2}{2|y_i|r_i}\right)
\end{align}
where $m(x)=(x+\pi\!\mod 2\pi)-\pi$ maps $x$ to the interval $[-\pi,\pi)$.

For the second \gls{sic}-stage of phase detection, we use algorithm 1 in \cite{jaeger:24:sic} with inputs $\vek{y}$, $\vek{r}$ and $\vek{\alpha}$. This algorithm uses approximate message passing so that $5\nsym-3$ messages, each containing a mean and variance, are passed along a factor graph to produce $\nsym$ outputs, each containing a mean and a variance. We collect the outputs in the vectors $\vek{\mu}$ and $\vek{\sigma}^2$, respectively. The corresponding a-posteriori metric for even $i$ is now \cite{jaeger:24:sic}
\begin{align}
    & q(\gamma_i|\vek{y},\vek{r},\vek{\alpha}) = \frac{h_i(\gamma_i)}{\sum_{\tilde{\gamma}}h_i(\tilde{\gamma})}\\
    & h_i(\gamma_i) = \mathcal{N}\left(m(\angle_{y_i}-\gamma_i-\mu_i);0,\sigma_i^2+\frac{\sigma_n^2}{2|y_i|r_i}\right).
\end{align}
For more \gls{sic}-stages, the algorithm described for the second stage is applied repeatedly. 

\section{Simulation Results}
We use the same simulation setup as in \cite{Secondini:19:Wiener,Gomez:20:CPAN,jaeger:24:sic} with a \SI{1000}{\kilo\meter} link and ideal distributed Raman amplification.
We use 24 sequences of 8192 symbols each to obtain $\incmean$, $\incvar$, $\thetavar$ and $\sigma_n^2$, and 120 sequences of 8192 symbols each for testing.
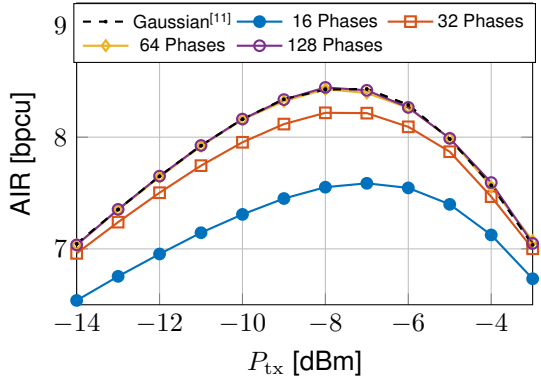
\begin{figure}
    \centering
    \begin{tikzpicture}
        \begin{axis}[%
        width=6cm,
        height=4cm,
        at={(16cm,0)},
        scale only axis,
        xmin=-14,
        xmax=-3,
        ymin=6.5,
        ymax=9.2,
        ytick = {7,8,9},
        xtick = {-14,-12,-10,-8,-6,-4},
        xlabel = {$P_\mathrm{tx}$ [dBm]},
        ylabel = {AIR [bpcu]},
        grid = major,
        axis background/.style={fill=white},
        legend pos=north west,
        legend style={nodes={scale=0.68, transform shape},at={(0.5,1)},anchor=north},
        legend columns = 3
        ]
            \addplot [color=black, dashed, style=thick, mark = *, mark size = .3pt]
              table[]{data/Ring_AIR_2_Stages-1.tsv};
              \addlegendentry{Gaussian \cite{jaeger:24:sic}};
            \addplot [color=matlab1, style=thick, mark = *]
              table[]{data/Discrete_Ring_AIR_2_Stages-1.tsv};
              \addlegendentry{16 Phases};
            \addplot [color=matlab2, style=thick, mark = square]
              table[]{data/Discrete_Ring_AIR_2_Stages-2.tsv};
              \addlegendentry{32 Phases};
            \addplot [color=matlab3, style=thick, mark = diamond]
              table[]{data/Discrete_Ring_AIR_2_Stages-3.tsv};
              \addlegendentry{64 Phases};
            \addplot [color=matlab4, style=thick, mark = o]
              table[]{data/Discrete_Ring_AIR_2_Stages-4.tsv};
              \addlegendentry{128 Phases};
            \addplot [color=black, dashed, forget plot, style=thick, mark = *, mark size = .3pt]
            table[]{data/Ring_AIR_2_Stages-1.tsv};
        \end{axis}
    \end{tikzpicture}
    \caption{\gls{air} of star-\gls{qam} with 32 rings and varying $\nphases$ compared to that of a \gls{cscg} using two \gls{sic}-stages.}
    \label{fig:air_2_stages}
\end{figure}

\figref{fig:air_2_stages} shows the \glspl{air} of star-\gls{qam} with 32 rings and varying cardinality of the phase constellation $\nphases$ using 2 \gls{sic}-stages. We also plot the \glspl{air} of \gls{cscg} modulation. The \gls{air} increases with $\nphases$ up to $\nphases=64$ where the \gls{air} of \gls{cscg} is met and saturates. Since the star-\gls{qam} constellations resemble \glspl{cscg} for a large number of rings and large $\nphases$, this result is intuitive. However, it is unclear if the loss for small $\nphases$ is due to poor constellation design or approximation errors arising from assuming $q(y_i|\theta_i)$ is constant in $\theta_i$. We remark that, since the \gls{air} of the phase channel is below \SI{5}{bpcu} \cite{jaeger:24:sic}, it could be supported by 32 phases. 

\begin{figure}
    \centering
    \begin{tikzpicture}
      \begin{axis}[%
                width=6.7cm,
                height=2.8cm,
                at = {(0cm,0)},
                scale only axis,
                xmin=10,
                xmax=36,
                xtick = {10,20,30},
                ymin=4,
                ymax=11.5,
                yticklabel = \empty,
                xlabel = {SNR [dB]},
                axis x line*=top,
                axis background/.style={fill=white},
                ]
                \addplot [color=matlabBlue, forget plot]
                  table[]{data/Discrete_Ring_AWGN-2.tsv};
                \addplot [color=matlabBlue, forget plot]
                  table[]{data/Discrete_Ring_AWGN-3.tsv};
                \addplot [color=matlabBlue, forget plot]
                  table[]{data/Discrete_Ring_AWGN-4.tsv};
                \addplot [color=matlabBlue, forget plot]
                  table[]{data/Discrete_Ring_AWGN-5.tsv};
                \addplot [color=matlabBlue, forget plot]
                  table[]{data/Discrete_Ring_AWGN-6.tsv};
                \addplot [color=matlabBlue, forget plot]
                  table[]{data/Discrete_Ring_AWGN-7.tsv};
                \addplot [color=matlabBlue, forget plot]
                  table[]{data/Discrete_Ring_AWGN-8.tsv};
                \addplot [color=white!15!black, forget plot]
                  table[]{data/Ring_AWGN-6.tsv};
                \addplot [color=matlabRed, forget plot]
                  table[]{data/Discrete_Ring_AWGN-1.tsv};
            \end{axis}
            \begin{axis}[%
                width=6.7cm,
                height=2.8cm,
                at = {(0cm,0)},
                scale only axis,
                xmin=-25.3004,
                xmax=0.6996,
                xtick = {-15,-5,5},
                ymin=4,
                ymax=11.5,
                xlabel = {$P_\mathrm{tx}$ [dBm]},
                axis x line* = bottom,
                grid = {major},
                legend pos=north west,
                legend style={nodes={scale=0.8, transform shape},at={(0,1)},anchor=north west},
                ]
                \addplot [draw=none, forget plot]
                  table[]{data/Ring_AWGN_Ptx-1.tsv};
                \addplot [color=matlabRed, draw = none]
                  table[]{data/Discrete_Ring_AWGN-1.tsv};
                \addlegendentry{Gaussian};
                \addplot [color=matlabBlue, draw = none]
                  table[]{data/Discrete_Ring_AWGN-2.tsv};
                  \addlegendentry{Star-\gls{qam}};
                \node[fill=white,text=matlabBlue, inner sep = 0pt] at (-0.5,5.3) {\small$2$};
                \node[fill=white,text=matlabBlue, inner sep = 0pt] at (-0.5,6.3) {\small$4$};
                \node[fill=white,text=matlabBlue, inner sep = 0pt] at (-0.5,7.3) {\small$8$};
                \node[fill=white,text=matlabBlue, inner sep = 0pt] at (-0.5,8.3) {\small$16$};
                \node[fill=white,text=matlabBlue, inner sep = 0pt] at (-0.5,9.3) {\small$32$};
                \node[fill=white,text=matlabBlue, inner sep = 0pt] at (-0.5,10.15) {\small$64$};
                \node[fill=white,text=matlabBlue, inner sep = 0pt] at (-0.5,10.6) {\small$128$};
            \end{axis}
        \end{tikzpicture}
    \caption{\gls{air} of memoryless AWGN channels with noise variance $\sigma_n^2$ %
    using 32 rings.\vspace{0.5cm}}
    \label{fig:memoryless_awgn}
\end{figure}
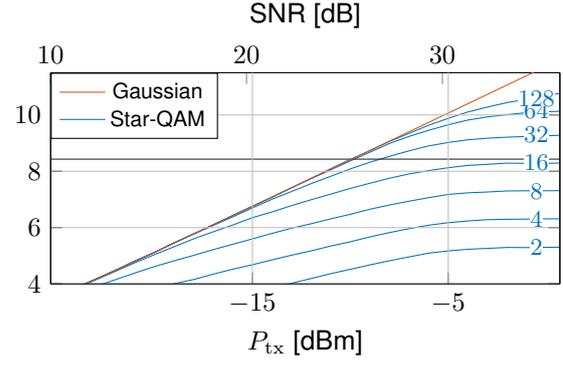

To investigate the issue, \figref{fig:memoryless_awgn} shows that star-\gls{qam} constellations with 32 rings and 32 phases experience a noticeable loss of \gls{air} compared to \gls{cscg} modulation even for memoryless \gls{awgn}-channels. We thus expect that the loss in \figref{fig:air_2_stages} is due to the small phase constellation cardinality rather than an approximation error. 

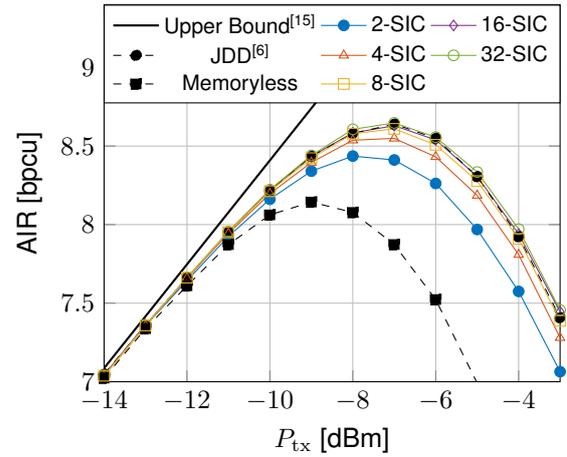
\begin{figure}[t]
    \centering
    \begin{tikzpicture}
            \begin{axis}[%
                width=6cm,
                height=5cm,
                at={(0,0cm)},
                scale only axis,
                xmin=-14,
                xmax=-3,
                ymin=7,
                ymax=9.4,
                ylabel = {AIR [bpcu]},
                xtick = {-14,-12,-10,-8,-6,-4},
                xlabel = {$P_\mathrm{tx}$ [dBm]},
                grid = major,
                axis background/.style={fill=white},
                legend style={nodes={scale=0.8, transform shape},at={(0.5,1)},anchor=north},
                legend columns = 3
                ]
                    
                  \addplot [color=black, style=thick]
                      table[]{data/air_upper_bound_ase-1.tsv};
                      \addlegendentry{Upper Bound \cite{Kramer:15:UpperBound}};
                    \addplot [color=matlab1, mark = *, mark size = 2pt]
                      table[]{data/AIR_32_Rings_128_Phases-1.tsv};
                    \addlegendentry{2-SIC};
                    \addplot [color=matlab4, mark = diamond, mark size = 2pt]
                      table[]{data/AIR_32_Rings_128_Phases-4.tsv};
                      \addlegendentry{16-SIC};
                    \addplot [dashed, color=black, mark = *, mark size = 2pt]
                      table[]{data/Gaussian_AIR-9.tsv};
                    \addlegendentry{JDD \cite{Gomez:20:CPAN}};
                    \addplot [color=matlab2, mark = triangle, mark size = 2pt]
                      table[]{data/AIR_32_Rings_128_Phases-2.tsv};
                      \addlegendentry{4-SIC};
                    \addplot [color=matlab5, mark = o, mark size = 2pt]
                      table[]{data/AIR_32_Rings_128_Phases-5.tsv};
                      \addlegendentry{32-SIC};
                      \addplot [dashed, color=black, mark = square*, mark size = 2pt]
                      table[]{data/memoryless_lb_discrete-1.tsv};
                    \addlegendentry{Memoryless};
                    \addplot [color=matlab3, mark = square, mark size = 2pt]
                      table[]{data/AIR_32_Rings_128_Phases-3.tsv};
                      \addlegendentry{8-SIC};
            \end{axis}
        \end{tikzpicture}
        \caption{\gls{air} of star-\gls{qam} with 32 rings, 128 phases, and a varying number of \gls{sic}-stages.}
        \label{fig:air_star_qam}
\end{figure}

\figref{fig:air_star_qam} plots the average \glspl{air} of star-\gls{qam} with 32 rings and 128 phases. We compare performance to those of a mismatched receiver based on memoryless \gls{awgn} surrogate channels, a receiver using \gls{jdd} and particle filtering \cite{Gomez:20:CPAN}, and an upper bound on capacity \cite{Kramer:15:UpperBound}. The curve for 2 \gls{sic}-stages is already closer to the \gls{jdd} curve than to the memoryless \gls{awgn} curve. This is because the \gls{awgn} curve does not account for the phase noise process and is smaller than the first-stage \gls{air}, and the second-stage \gls{air} is already close to the \gls{jdd} \gls{air}. Observe that the average \gls{air} increases with the number of \gls{sic}-stages until it saturates near the \gls{air} of \gls{jdd} at 16 or more \gls{sic}-stages. The curves suggest that 2-4 \gls{sic}-stages might suffice for practical implementations.

\section{Outlook}
The proposed transmitter and receiver show competitive performance in mitigating phase noise with long correlations, similar to a genie-aided receiver that can fully compensate for phase noise \cite{jaeger:24:sic}. To further increase the \gls{air}, one might focus on the \gls{isi} caused by \gls{xpm}. 

For future work, the receiver could be generalized to incorporate single-channel \gls{dbp}. For this purpose, one must pay attention to nonlinear phase noise and strong additive noise terms caused by \gls{spm} \cite{Dar:13:NLIN}. Note that \gls{spm} can create additive noise through strong two-pulse collisions, whereas \gls{xpm} creates only phase noise noise through such collisions \cite{Dar:16:Collision}.

\section{Conclusion}
We extended the \gls{sic}-receiver studied in \cite{jaeger:24:sic} to discrete modulation alphabets, in particular probabilistically-shaped star-\gls{qam} constellations. We recovered the \glspl{air} computed in \cite{jaeger:24:sic} using \gls{sic} and \gls{cscg}, and the \glspl{air} in \cite{Gomez:20:CPAN} using \gls{jdd} and particle filtering, for a \SI{1000}{\kilo\meter} transmission link. We found that 32 rings and 128 phases suffice when using 16 \gls{sic}-stages. Also, 2-4 \gls{sic}-stages offer a good trade-off between performance and computational cost and might hence be interesting for practical implementations.

\clearpage
\section{Acknowledgements}
The authors acknowledge the financial support by the Federal Ministry of Education and Research of Germany in the programme of “Souverän. Digital. Vernetzt.” Joint project 6G-life, project identification number: 16KISK002.

\defbibnote{myprenote}{%
}
\printbibliography[prenote=myprenote]

@ARTICLE{Gomez:20:CPAN,
  author={García-Gómez, Francisco Javier and Kramer, Gerhard},
  journal={Journal of Lightwave Technology}, 
  title={Mismatched Models to Lower Bound the Capacity of Optical Fiber Channels}, 
  year={2020},
  volume={38},
  number={24},
  pages={6779-6787},
  doi={10.1109/JLT.2020.3021277}}

@ARTICLE{Secondini:19:Wiener,
  author={Secondini, Marco and Agrell, Erik and Forestieri, Enrico and Marsella, Domenico and Camara, Menelaos Ralli},
  journal={Journal of Lightwave Technology}, 
  title={Nonlinearity Mitigation in WDM Systems: Models, Strategies, and Achievable Rates}, 
  year={2019},
  volume={37},
  number={10},
  pages={2270-2283},
  doi={10.1109/JLT.2019.2901908}}

@ARTICLE{Colavolpe:05:Iterative,
  author={Colavolpe, G. and Barbieri, A. and Caire, G.},
  journal={IEEE Journal on Selected Areas in Communications}, 
  title={Algorithms for iterative decoding in the presence of strong phase noise}, 
  year={2005},
  volume={23},
  number={9},
  pages={1748-1757},
  doi={10.1109/JSAC.2005.853813}}

@ARTICLE{Plabst:24:SIC,
  author={Plabst, Daniel and Prinz, Tobias and Diedolo, Francesca and Wiegart, Thomas and B\"ocherer, Georg and Hanik, Nobert and Kramer, Gerhard},
  journal={IEEE Transactions on Communications, submitted}, 
  title={Neural network equalizers and successive interference cancellation for bandlimited channels with a nonlinearity}, 
  year={2024},
  archivePrefix={arXiv},
  eprint={2401.09217},}

@article{Wachsmann:99:Multilevel,
  author={Wachsmann, U. and Fischer, R.F.H. and Huber, J.B.},
  journal={IEEE Transactions on Information Theory}, 
  title={Multilevel codes: theoretical concepts and practical design rules}, 
  year={1999},
  volume={45},
  number={5},
  pages={1361-1391},
  doi={10.1109/18.771140}}

@INPROCEEDINGS{Pfister:01:ISI,
  author={Pfister, H.D. and Soriaga, J.B. and Siegel, P.H.},
  booktitle={GLOBECOM'01. IEEE Global Telecommunications Conference}, 
  title={On the achievable information rates of finite state {ISI} channels}, 
  year={2001},
  volume={5},
  number={},
  pages={2992-2996},
  doi={10.1109/GLOCOM.2001.965976}}

@ARTICLE{Mecozzi:12:RP,
  author={Mecozzi, Antonio and Essiambre, René-Jean},
  journal={Journal of Lightwave Technology}, 
  title={Nonlinear Shannon Limit in Pseudolinear Coherent Systems}, 
  year={2012},
  volume={30},
  number={12},
  pages={2011-2024},
  doi={10.1109/JLT.2012.2190582}}

@article{Dar:13:NLIN,
    author = {Ronen Dar and Meir Feder and Antonio Mecozzi and Mark Shtaif},
    journal = {Optics Express},
    number = {22},
    pages = {25685--25699},
    title = {Properties of nonlinear noise in long, dispersion-uncompensated fiber links},
    volume = {21},
    year = {2013},
    doi = {10.1364/OE.21.025685},
}

@ARTICLE{Dar:16:Collision,
  author={Dar, Ronen and Feder, Meir and Mecozzi, Antonio and Shtaif, Mark},
  journal={Journal of Lightwave Technology}, 
  title={Pulse Collision Picture of Inter-Channel Nonlinear Interference in Fiber-Optic Communications}, 
  year={2016},
  volume={34},
  number={2},
  pages={593-607},
  doi={10.1109/JLT.2015.2428283}}

@ARTICLE{Yankov:15:Phase,
  author={Yankov, Metodi P. and Fehenberger, Tobias and Barletta, Luca and Hanik, Norbert},
  journal={Journal of Lightwave Technology}, 
  title={Low-Complexity Tracking of Laser and Nonlinear Phase Noise in WDM Optical Fiber Systems}, 
  year={2015},
  volume={33},
  number={23},
  pages={4975-4984},
  doi={10.1109/JLT.2015.2493202}}

@INPROCEEDINGS{Kramer:15:UpperBound,
  author={Kramer, Gerhard and Yousefi, Mansoor I. and Kschischang, Frank R.},
  booktitle={2015 IEEE Information Theory Workshop (ITW)}, 
  title={Upper bound on the capacity of a cascade of nonlinear and noisy channels}, 
  year={2015},
  volume={},
  number={},
  pages={1-4},
  doi={10.1109/ITW.2015.7133167}}

@article{ten2004design,
  author={ten Brink, S. and Kramer, G. and Ashikhmin, A.},
  journal={IEEE Transactions on Communications}, 
  title={Design of low-density parity-check codes for modulation and detection}, 
  year={2004},
  volume={52},
  number={4},
  pages={670-678},
  doi={10.1109/TCOMM.2004.826370}}

@article{jaeger:24:sic,
    title={Information Rates of Successive Interference Cancellation for Optical Fiber}, 
    author={Alex Jäger and Gerhard Kramer},
    year={2024},
    eprint={2403.15240},
    journal={IEEE Journal on Selected Areas in Communications, submitted},
    archivePrefix={arXiv}}

@ARTICLE{Vannucci:02:RP,
  author={Vannucci, A. and Serena, P. and Bononi, A.},
  journal={Journal of Lightwave Technology}, 
  title={The RP method: a new tool for the iterative solution of the nonlinear Schrodinger equation}, 
  year={2002},
  volume={20},
  number={7},
  pages={1102-1112},
  doi={10.1109/JLT.2002.800376}}

@article{douillard1995iterative,
  title={Iterative correction of intersymbol interference: turbo-equalization},
  author={Douillard, Catherine and J{\'e}z{\'e}quel, Michel and Berrou, Claude and Electronique, D{\'e}partement and Picart, Annie and Didier, Pierre and Glavieux, Alain},
  journal={European Transactions Telecommunications},
  volume={6},
  number={5},
  pages={507-511},
  year={1995},
  doi = {10.1002/ett.4460060506}
}

\vspace{-4mm}

\end{document}